\documentclass[pre,twocolumn,eqsecnum,showpacs]{revtex4}
\usepackage{amssymb,bm}
\usepackage[dvips]{graphicx}
\usepackage{psfrag}

\begin{document}

\title{Magnetic field correlations in a random flow with strong steady shear}

\author{I. Kolokolov, V. Lebedev, and G. Sizov}

\affiliation{Landau Institute for Theoretical Physics RAS, \\
 119334, Kosygina 2, Moscow, Russia, \\
 and Moscow Institute of Physics and Technology.}

\date{\today}

\begin{abstract}

We analyze magnetic kinematic dynamo in a conducting fluid where the stationary shear
flow is accompanied by relatively weak random velocity fluctuations. The diffusionless
and diffusion regimes are described. The growth rates of the magnetic field moments are
related to the statistical characteristics of the flow describing divergence of the
Lagrangian trajectories. The magnetic field correlation functions are examined, we
establish their growth rates and scaling behavior. General assertions are illustrated by
explicit solution of the model where the velocity field is short-correlated in time.

\end{abstract}


\maketitle

 \section{Introduction}
 \label{sec:intro}

The subject of our work is the magnetic dynamo that is the magnetic field generation by
hydrodynamic motions in a conducting medium. We theoretically investigate the effect in a
conducting fluid (plasma, electrolyte) where a random hydrodynamic flow is excited. The
principal example of such flow is hydrodynamic turbulence (see, e.g., Refs.
\cite{MY,Frisch}) responsible for the magnetic field generation in different geophysical
and astrophysical phenomena, see Refs. \cite{VZ72,MHD,P79,ZRS,CG95,astro1,astro2,astro3}.
We consider the case where the magnetic field grows from small initial fluctuations and
examine the evolution stage when the magnetic field is weak enough, so that one can
neglect feedback from the magnetic field to the flow. The stage where the flow is
independent of the magnetic field is called kinematic. The kinematic approach becomes
invalid when the increasing magnetic field begins to affect the fluid motions
essentially. In this case the velocity field is strongly influenced by the Lorentz force,
so that the induction dynamics is no longer linear. In most cases that leads to
saturation of the magnetic field fluctuations maintained by the hydrodynamic flow. Though
the magnetic field cannot be described by a linear equation in the regime, the kinematic
stage produces magnetic structures similar to those occurring at the saturation state,
see, e.g., Ref. \cite{schek}. A possible explanation of the fact is related to strong
intermittency of the magnetic field that implies that the feedback is concentrated in
restricted space regions where the magnetic field is anomalously strong.

We assume that the random flow, exciting dynamo, is statistically homogeneous in space
and time. Usually an additional assumption is made that the flow is statistically
isotropic. If the velocity field is short-correlated in time then it is possible to
derive closed equations for the magnetic induction correlation functions (see, e.g., Ref.
\cite{01FGV}). The pair correlation function for the case has been analyzed in Refs.
\cite{Kra,Kaz}. The complete statistical description of the magnetic field for a
short-correlated smooth statistically isotropic flow has been done in Ref. \cite{99CFKV}
where growth rates and structure of spatial correlation functions were found. However, it
is of interest to consider random flows with an average shear flow, that are widespread
in astrophysical applications. Such flows are statistically anisotropic and need a
special analysis. Here we examine the case where a steady shear flow is complemented by a
relatively weak random component. We focus on the analysis of growth rates of moments of
the magnetic field (magnetic induction), the degree of its anisotropy and on structure of
the magnetic field correlation functions. We aim to relate the magnetic statistical
characteristics to those of the flow, thus revealing the most universal features of the
dynamo effect. The general assertions are illustrated by the model where the velocity
field is short correlated in time, that can be solved analytically.

An additional motivation for our research comes from dynamics of polymer solutions that
in many respects is similar to magneto-hydrodynamics, see Refs. \cite{Bird,Doi}.
Particularly, we have in mind the so-called coil-stretch transition \cite{Lumley} (see
also Refs. \cite{00BFL,05CKLT}) that is an analog of the dynamo effect. A decade ago the
elastic turbulence was discovered (see Refs. \cite{00GS,01GSa,01GSb}) that is a chaotic
hydrodynamic motion of polymer solutions, the elastic turbulence can be realized even at
small Reynolds numbers as against the traditional hydrodynamic turbulence. The elastic
turbulence is a natural frame for applying an extension of the dynamo theory to polymer
solutions.

The behavior of the magnetic field moments at the kinematic stage in the presence of the
strong shear flow was established in Ref. \cite{10KKL}. However, to examine the spacial
structure of the magnetic field, one should know its correlation functions, that are
studied in the present work in the framework of the same general scheme as in Ref.
\cite{10KKL}. To verify our general predictions we examine the analytically solvable
model where a short-correlated in time random flow is excited on background of the strong
stationary shear flow.

The structure of our paper is as follows. In Section \ref{sec:basic} we introduce basic
relations needed to analyze the magnetic field correlations and dynamics. We present
general dynamic equation, give its formally exact solution and discuss statistical
properties of quantities entering this solution. In Section \ref{sec:corre} moments and
correlation functions of the magnetic field are investigated. We relate its growth rates
to the growth rates of the separation between two close fluid particles and establish
principal space structure of the correlation functions. Section \ref{sec:short} is
devoted to the model where fluctuating component of the flow is short correlated in time.
We establish the growth rates for the model and analyze in detail the pair correlation
function. The obtained results are in agreement with our general assertions. In Section
\ref{sec:discussion}  we outline our main results and discuss their possible applications
and extensions.

 \section{Basic relations}
 \label{sec:basic}

We consider the magnetic field evolution in a conducting fluid (plasma or electrolyte)
where hydrodynamic motions are excited. Then the magnetic field dynamics is governed by
the following equation (see, e.g., Ref. \cite{LL})
 \begin{equation}
 \partial_t{\bm B}=
 ({\bm B}\cdot{\bm\nabla}){\bm v}
 -({\bm v}\cdot{\bm\nabla}){\bm B}
 + \kappa\nabla^2{\bm B},
 \label{magnet}
 \end{equation}
where $\bm B$ is the magnetic induction, $\bm v$ is the flow velocity and $\kappa$ is the
magneto-diffusion coefficient, inversely proportional to the electrical conductivity of
the medium. The flow is assumed to be incompressible, $\nabla\cdot\bm v=0$. We also
assume that the magneto-diffusion term in Eq. (\ref{magnet}) is small in comparison with
those related to the flow. We consider the case where the magnetic field is relatively
weak and, therefore, its feedback to the flow is negligible. Then the relation
(\ref{magnet}) is a linear equation determining the magnetic field evolution in a
prescribed velocity field, this regime is called kinematic.

The hydrodynamic motion excited in the fluid is assumed to be random (turbulent) and the
velocity statistics is assumed to be homogeneous in space and time. We examine the
magnetic field growth from initial weak fluctuations distributed statistically
homogeneously in space at the initial time $t=0$. The correlation length of the initial
fluctuations $l$ is assumed to be smaller than the velocity correlation length $\eta$. If
we consider hydrodynamic turbulence then a role of the velocity correlation length is
played by the Kolmogorov scale. At scales less than $\eta$ the velocity field $\bm v$ can
be treated as smooth. The magnetic growth (dynamo) can be characterized by moments of the
magnetic induction that exponentially increase over time $t$:
 \begin{equation}
 \left\langle \left|\bm B(t)\right|^{2n} \right\rangle
 \propto\exp\left(\gamma_n t\right),
 \label{increments}
 \end{equation}
Here angular brackets mean averaging over space. The exponential laws (\ref{increments})
are characteristic of the kinematic dynamo since the equation (\ref{magnet}) is linear in
the magnetic induction $\bm B$ in this case.

One of our goal is to express the growth rates (increments) $\gamma_n$ in Eq.
(\ref{increments}) via statistical characteristics of the flow. The natural measure for
the growth rates $\gamma_n$ is the so-called Lyapunov exponent of the flow, $\lambda$,
that is equal to the average logarithmic divergence rate of close fluid particles. A
special question concerns an $n$-dependence of $\gamma_n$. If the magnetic induction
statistics is Gaussian then $\gamma_n\propto n$. Deviations from the linear law signal
about intermittency of the magnetic field. The intermittency implies that high moments of
the magnetic field are determined by rare strong fluctuations.

There are two different regimes of the kinematic magnetic field growth. The first regime
is realized if all characteristic scales of the magnetic field are much larger than the
magnetic diffusion length $r_d$, $r_d=\sqrt{\kappa/\lambda}$. The assumed smallness of
the diffusion coefficient implies the inequality $\eta\gg r_d$. We assume also $l\gg
r_d$, then the diffusion term in Eq. (\ref{magnet}) is negligible at the first stage of
the magnetic evolution that we call diffusionless. The magnetic force lines are deformed
by the flow without reconnections in this regime. However, distortions of the magnetic
field by the flow inevitably lead to producing scales of order $r_d$ in the field. After
that the magnetic diffusion is switched on that admits reconnections. This second
(diffusion) stage is characterized by the growth rates different from ones describing the
diffusionless regime.

Let us present a qualitative picture explaining the magnetic field evolution at the
kinematic stage. The initial magnetic field distribution in space can be thought as an
ensemble of blobs of sizes $\sim l$. Then the blobs are distorted by the flow being
stretched in one direction and compressed in another direction. In the isotropic case
directions of the stretching and compression vary chaotically in space and time whereas
in our case they are attached to the shear flow: the blobs are stretched mainly along the
shear velocity and are compressed in the direction of the shear velocity gradient. At the
first (diffusionless) stage the blobs are deformed without intersections and the magnetic
field induction grows like a separation between close fluid particles since the equation
(\ref{magnet}) at $\kappa=0$ coincides with the equation for the separations.

\begin{figure}[t]
 \includegraphics[width=3.5in,angle=0]{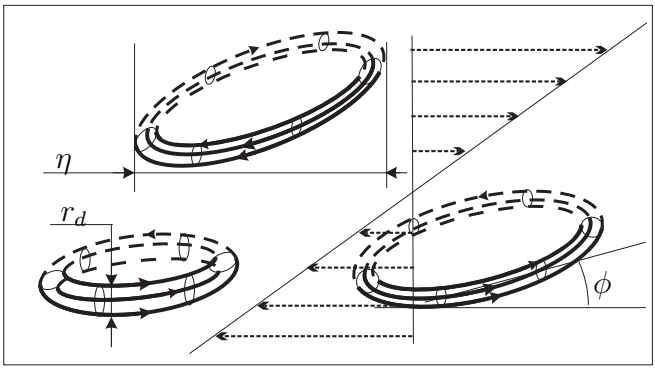}
 \caption{Sketch of typical magnetic blobs during the diffusive kinematic stage.}
 \label{fig:blobs}
 \end{figure}

The diffusionless stage finishes when the characteristic blob width diminishes down to
the diffusion length $r_d$. Then the diffusion is switched on that leads to two effects.
First, the diffusion prevents further shrinking the blob widths, so that they remain of
the order of $r_d$, whereas the blobs continue to be stretched in the direction of the
shear velocity. Second, due to reconnections of the magnetic force lines admitted by
diffusion the blobs start to overlap. As a result, new blobs of a characteristic
longitudinal size $\eta$ are formed, see Fig.\ref{fig:blobs}. The magnetic induction in
such blobs can be found by averaging the induction of a large number $N$ of initial
blobs, the number $N$ grows exponentially as time runs. Averaging over the large number
of random variables leads to appearing an exponentially small factor $\sim 1/\sqrt N$ in
the amplitude of the magnetic induction. Besides, the amplitudes of the initial blobs
remain to increase over time as a separation between fluid particles. We conclude that at
this second (diffusive) stage the magnetic field is still growing exponentially over time
but slower than at the first stage.

 \psfrag{Theta}{\kern0pt\lower0pt  \hbox{\large$\phi$}}
 \psfrag{Eta}{\kern0pt\lower0pt  \hbox{\large$\eta$}}
 \psfrag{Rd}{\kern0pt\lower0pt  \hbox{\large$r_d$}}

We consider the case where the steady shear constituent of the flow is much stronger than
the random one. Quantitatively, the condition is written as the inequality $s\gg\lambda$
where $s$ is the shear rate. Indeed, the Lyapunov exponent in a pure shear flow is zero,
and its non-zero value is associated with presence of the relatively weak random
constituent of the flow. The distorted magnetic blobs are elongated mainly along the
shear velocity. However, they are tilted with respect to the velocity direction due to
presence of the random velocity component, see Fig. \ref{fig:blobs}. The tilt possesses
the same dynamics as the direction of the polymer stretching in the same flow, see Ref.
\cite{05CKLT}. Therefore the tilt angle $\phi$, see Fig. \ref{fig:blobs}, can be
estimated as $\phi\sim\lambda/s$. The tilt angle determines typical ratio of the magnetic
field components $B_y/B_x\sim\lambda/s\ll1$ where the $X$-axis is directed along the
shear velocity that varies along the $Y$-axis. Thus, the ratio $s/\lambda$ characterizes
an anisotropy degree of the magnetic field.

 \subsection{Lagrangian dynamics}

To analyze moments and correlation functions of the magnetic induction we need a solution
of the magneto-dynamic equation (\ref{magnet}) for the induction field $\bm B(t)$ in
terms of its initial value ${\cal B}$, ${\cal B}=\bm B(0)$. We use here a generalization
of the scheme proposed in Ref. \cite{97CKV} that exploits the Lagrangian approach to the
fluid motion. Passing to the Lagrangian frame, that moves with the fluid particles, one
excludes the advection term in Eq. (\ref{magnet}). Next, instead of solving the equation
with the Laplace operator $\nabla^2$ it is convenient to introduce Langevin forces
$\bm\xi$ mimicing the magnetic diffusion. Then we pass to the following stochastic
equation
 \begin{equation}
 \partial_t \bm R = \bm v(t,\bm R)+ \bm\xi,
 \label{Lagrange}
 \end{equation}
describing Lagrangian trajectories disturbed by the Langevin forces. The latter can be
treated as white noise characterized by its pair correlation function
 \begin{equation}
 \langle \xi_i(t_1) \xi_j(t_2) \rangle
 =2\kappa \delta_{ij}\delta (t_1-t_2),
 \label{white}
 \end{equation}
where $\kappa$ is the diffusion coefficient figuring in Eq. (\ref{magnet}).

Rewriting the equation (\ref{magnet}) in terms of the quantity $\bm R(t)$ one finds then
the following formally exact solution of Eq. (\ref{magnet})
 \begin{equation}
 \bm B(t,\bm r)= \left\lfloor\hat W(t) {\cal B}[\bm R(0)]\right\rfloor,
 \label{solution}
 \end{equation}
where floors mean averaging over the $\bm\xi$-statistics, in accordance with Eq.
(\ref{white}). To find the quantity $\bm R(0)$ one should solve the equation
(\ref{Lagrange}) on the time interval $(0,t)$ with the boundary condition $\bm R(t)=\bm
r$, posed at the final time. By other words, one should track the magnetic field back in
time along the disturbed Lagrangian trajectories. The matrix $\hat W(t)$ in Eq.
(\ref{solution}) is a chronologically ordered exponent
 \begin{equation}
 \hat W(t) = \mathrm T \exp\left\{
 \int_0^t dt'\, \hat\Sigma (t') \right\},
 \label{evolution}
 \end{equation}
where $\hat\Sigma(t)$ is the velocity gradients matrix, $\Sigma_{ji}=\partial_i v_j$,
taken at the time $t$ and in the spacial point $\bm R(t)$. The matrix $\hat W$, that we
call an evolution matrix, can be treated as a solution of the equation $\partial_t\hat
W=\hat\Sigma \hat W$ with the initial condition $\hat W(0)=1$.

The evolution matrix $\hat W$ has some general properties that follow from the definition
(\ref{evolution}). The determinant of the matrix $\hat W$ is equal to unity, since the
velocity gradient matrix $\hat\Sigma$ is traceless, $\mathrm{tr}\, \hat\Sigma=0$, that,
in turn, is a consequence of the incompressibility condition $\nabla\cdot\bm v=0$.
Therefore $W_1 W_2 W_3=1$, where $W_1,W_2,W_3$ are eigen values of the matrix $\hat W$.
All the eigen values are positive since they are positive initially ($W_1=W_2=W_3=1$ at
$t=0$) and cannot turn to zero because of the same relation $W_1 W_2 W_3=1$. Let us order
the eigen values in accordance with $W_1>W_2>W_3$, then $W_1>1$ and $W_3<1$. At times
$t\gg \lambda^{-1}$, we are interested in, typical values of $\ln W_1, \ln W_3$ can be
estimated as $\pm\lambda t$, therefore $W_1$ is exponentially large whereas $W_3$ is
exponentially small. An estimation for $W_2$ depends on details of the flow statistics.
Any case, $W_1\gg W_2 \gg W_3$ at times $t\gg \lambda^{-1}$.

In the framework of the proposed formalism, correlation functions of the magnetic field
$\bm B$ have to be calculated by averaging products of the factors (\ref{solution}) taken
at the respective points over the statistics of the noises $\bm\xi$, besides averaging
over space. Thus, say, the simultaneous correlation function,
 \begin{eqnarray}
 F_{2n,i \dots j}(\bm r_1, \dots, \bm r_{2n})
 =\left\langle B_i(\bm r_1) \dots
 B_j(\bm r_{2n}) \right\rangle,
 \label{corrfun}
 \end{eqnarray}
has to be calculated in two steps. First, one should substitute the expression
(\ref{solution}) into the right-hand side of Eq. (\ref{corrfun}) and then average the
resulting product over the $\xi$-statistics given by Eq. (\ref{white}), this averaging
catches diffusion effects. Let us underline that the fields $\bm\xi$ have to be treated
as independent for all $2n$ factors in the product. Second, one should average the result
over space. An averaging over scales $\lesssim \eta$ (traced back to the initial time)
gives statistics of the initial magnetic field fluctuations, and averaging over scales
$\gtrsim \eta$ counts different realizations of $\hat\Sigma$. Therefore the latter is
equivalent to averaging over the velocity statistics. This logic was realized for the
isotropic random flow in Ref. \cite{99CFKV}.

In the diffusionless regime, realized at $t\ll \lambda^{-1} \ln(l/r_d)$, one can neglect
diffusion effects. Then at calculating the moment $\langle|\bm B|^{2n}\rangle$ one can
take a product of the identical factors (\ref{solution}) where $\bm R$ is simply a
Lagrangian trajectory terminated at the point $\bm r$ at time $t$. Then $|\bm B(\bm
r)|^{2n}\approx W_1^{2n} |{\cal B}|^{2n}$ where ${\cal B}$ is taken at the origin of the
Lagrangian trajectory. Here just the factor $W_1^{2n}$ is responsible for the exponential
growth of the moments, and therefore, we can restrict ourselves to the estimation
$\lfloor|\bm B(\bm r)|^{2n}\rfloor \sim W_1^{2n} {\cal B}_0^{2n}$ where ${\cal B}_0$ is
the characteristic value of the initial magnetic field fluctuations. In the diffusion
regime, realized at $t\gg \lambda^{-1} \ln(l/r_d)$, the situation is a bit more
complicated.

Let us first consider the second moment. Then we deal with two trajectories, $\bm R$ and
$\bm R'$, terminating at the same point $\bm r$ at time $t$, but characterized by
independent noises $\bm\xi$ and $\bm\xi'$. The second moment can be written as the
following average
 \begin{equation}
 \langle(\bm B)^2\rangle=
 \left\langle\left\lfloor {\cal B}^T[\bm R(0)] \hat W^T
 \hat W' {\cal B}[\bm R'(0)]\right\rfloor\right\rangle,
 \label{second}
 \end{equation}
where the subscript $T$ means transposition of the object. An appreciable contribution to
the second moment is related to the trajectories with $|\bm R(0)-\bm R'(0)|\lesssim l$.
Since $|\bm R(0)-\bm R'(0)|\ll \eta$ and $|\bm R(t)-\bm R'(t)|=0$, the difference
$\Delta\bm R=\bm R-\bm R'$ stays to be much less than $\eta$ at any time from the
interval $(0,t)$ for such event. Then we obtain from Eq. (\ref{Lagrange}) expanding the
velocity up to linear in $\Delta\bm R$ terms
 \begin{equation}
 \partial_t\Delta\bm R=
 \hat\Sigma \Delta\bm R
 +\bm\xi-\bm\xi',
 \label{divergence}
 \end{equation}
where $\hat\Sigma$ can be taken at the point $\bm R$ or $\bm R'$, no difference. A
solution of Eq. (\ref{divergence}), equal to zero at time $t'=t$, is written as
 \begin{equation}
 \Delta\bm R(t')= -\hat W(t') \int_{t'}^t dt_1
 \hat W^{-1} (t_1)[\bm\xi(t_1)-\bm\xi'(t_1)].
 \label{difference}
 \end{equation}

To calculate the second moment we should know the $\Delta\bm R(0)$-statistics. Since the
separation $\Delta\bm R(0)$ is a linear combination of $\bm\xi,\bm\xi'$ it should be
treated as a Gaussian variable at averaging over the $\bm\xi$-statistics, then its
probability distribution function is completely determined by the matrix $\hat M$:
 \begin{equation}
 M_{ij}=\left\lfloor \Delta R_i(0) \Delta R_j(0) \right\rfloor
 =4\kappa \int_0^t dt_1\, W^{-1}_{ik} W^{-1}_{jk},
 \label{secondd}
 \end{equation}
the expression for $\hat M$ is derived from Eqs. (\ref{white},\ref{difference}). The
matrix $\hat M$ is symmetric, and its eigen values are positive. Let us designate the
eigen values as $m_1^2,m_2^2,m_3^2$ and order the values as $m_1>m_2>m_3$, the
inequalities become strong, $m_1\gg m_2\gg m_3$, if $\lambda t\gg 1$. Let us stress that
the directions of the eigen vectors of the matrix $\hat M$ are ``frozen'' at $\lambda
t\gg1$ \cite{Furst,Orszag,99BF}. Then the integral determining $m_1$ is gained at
$t-t_1\sim \lambda^{-1}$ and we arrive at the estimation $m_1 \sim r_d W_3^{-1}$. The
integral determining $m_3$ is gained at $t_1\sim \lambda^{-1}$ and therefore $m_3 \sim
r_d$. An estimation for $m_2$ depends on the time dependence of $W_2$. If $W_2$ increases
then $m_2$ remains of the order of $r_d$ whereas it grows like $m_2\sim r_d W_2^{-1}$ if
$W_2$ decreases.

Now we find a probability that $\Delta\bm R(0)$ is less than $l$ in the diffusion regime,
when $t\gg \lambda^{-1} \ln(l/r_d)$. One can think in terms of the components of
$\Delta\bm R(0)$ in the basis attached to the eigen vectors of the matrix $\hat M$. Since
$m_1\gg l$ then a probability that the first component of $\Delta\bm R(0)$ is less than
$l$ is estimated as $l/m_1\sim (l/r_d) W_3$. If $W_2$ grows as time runs then both, $m_2$
and $m_3$, are of the order of $r_d$ and therefore a probability that the second and the
third components of $\Delta\bm R(0)$ are less than $l$, is close to unity. Then we find
from Eq. (\ref{second})
 \begin{equation}
 \lfloor |\bm B|^2 \rfloor
 \sim {\cal B}_0^2(l/r_d){W_1}W_2^{-1} ,
 \label{diffusive}
 \end{equation}
where ${\cal B}_0$ is a characteristic value of the initial magnetic field fluctuations
and we used the relation $W_1 W_2 W_3=1$.

The situation with the decreasing $W_2$ is slightly different. In this case $m_2\gg l$ at
the diffusive stage and there appears an additional small probability that the second
component $\Delta\bm R(0)$ is less than $l$, the probability can be estimated as
$l/m_2\sim (l/r_d) W_2$. Then one would obtain  $\langle |\bm B|^2 \rangle \sim {\cal
B}_0^2{W_1} (l/r_d)^2$, instead of Eq. (\ref{diffusive}). However, an integration over
space (at the next step of averaging) kills the leading term due to the solenoidal nature
of the magnetic field $\bm B$. Therefore one has to take into account the next term in
the probability distribution of $\Delta R_2(0)$ that gives an extra small factor
$(l/m_2)^2$ in the probability. Thus, we arrive at
 \begin{equation}
 \lfloor |\bm B|^2 \rfloor \sim {\cal B}_0^2 (l/r_d)^4
 W_1 W_2^2.
 \label{diffusive1}
 \end{equation}

Note that the expressions (\ref{diffusive},\ref{diffusive1}) are equivalent to ones
obtained in the Fourier representation for the statistically isotropic case in Ref.
\cite{99CFKV}. However, the expressions (\ref{diffusive},\ref{diffusive1}), written for
real space, are correct for the anisotropic problem (we are investigating) as well, and
are in fact more suitable for the problem.

Let us turn to the high moments. One can recognize that a principal contribution to the
average $\lfloor |\bm B|^{2n} \rfloor$ is produced by configurations where the $2n$
points $\bm R_\alpha(0)$ are divided into $n$ pairs with separations $\lesssim l$ in each
pair. Because of the independence of the white noises $\bm\xi_\alpha$, the probability of
such event can be estimated as a product of probabilities for the second moment, that is
 \begin{equation}
 \lfloor |\bm B|^{2n} \rfloor \sim
 \lfloor |\bm B|^{2} \rfloor ^n ,
 \label{high}
 \end{equation}
where the second moment is given by Eq. (\ref{diffusive}) or Eq. (\ref{diffusive1}). We
have ignored a combinatoric factor in Eq. (\ref{high}) being interested in the time
dependence of the moments.

 \subsection{Evolution matrix}

The next step in finding the magnetic field moments is averaging over the velocity
statistics. Before doing it we should establish statistical properties of the evolution
matrix (\ref{evolution}). Some universal properties of such matrices, that can be treated
as products of a large number of random matrices, are well established, see Refs.
\cite{Furst,Orszag,99BF}, the properties are revealed at $t\gg \lambda^{-1}$. However, we
examine a strongly anisotropic case, with steady shear flow dominating. That forced us to
modify the consideration of Ref. \cite{99CFKV} where the isotropic case (statistically
isotropic flow) was investigated.

For the anisotropic problem, it is convenient to use the Gaussian decomposition of the
evolution matrix $\hat{W}=\hat{T}_L\hat{\Delta}\hat{T}_R$, where $\hat{T}_L$ and
$\hat{T}_R$ are triangle matrices,
 \begin{eqnarray}
 \hat{T}_L=\left(
 \begin{array}{ccc}
 1 & \chi & \chi_1 \\
 0 & 1 & \chi_2 \\
 0 & 0 & 1
 \end{array}\right), \quad
 \hat{T}_R=\left(
 \begin{array}{ccc}
 1 & 0 & 0 \\
 \zeta_1 & 1 & 0 \\
 \zeta_2 & \zeta_3 & 1
 \end{array}\right),
 \label{e6}
 \end{eqnarray}
and $\hat\Delta$ is a diagonal matrix. Since both triangle matrices, $\hat T_L$ and $\hat
T_R$, have unit determinants, the determinant of $\hat\Delta$ is equal to unity as well.

The matrices are written in the reference frame attached to the shear flow: the axis $X$
is directed along the shear velocity and the axis $Y$ is directed along the shear
velocity gradient. Therefore the shear velocity is written as $v_x=sy$ where $s$ is the
shear rate. For our flow, which is composed of the steady shear flow and a random
component, the matrix of the velocity gradients $\Sigma_{ji}=\partial_i v_j$ is a sum of
two terms related to the shear and the random components of the flow:
 \begin{equation}
 \Sigma_{ji}(t)=s\delta_{jx}\delta_{iy}+\sigma_{ji}(t).
 \label{sig}
 \end{equation}
The random matrix $\sigma_{ji}$ is zero in average and should be characterized in terms
of its correlation functions. The trace of the matrix is zero, $\mathrm{tr}\,
\hat\sigma=0$ (due to the flow incompressibility). Let us remind that the Lyapunov
exponent $\lambda$ of a purely shear flow is equal to zero. Therefore $\lambda$ is
sensitive to $\hat\sigma$ though the random flow is weaker than the steady one.

Substituting the decomposition $\hat{W}=\hat{T}_L\hat{\Delta}\hat{T}_R$ into the
evolution equation $\partial_t\hat W =\hat \Sigma \hat W$, one finds
 \begin{equation}
 \hat{T}_L^{-1}\hat{\Sigma}\hat{T}_L=
 \hat{T}_L^{-1}\partial_t{\hat{T}}_L
 +\partial_t{\hat{\Delta}}\hat{\Delta}^{-1}+
 \hat{\Delta}\partial_t{\hat{T}}_R\hat{T}_R^{-1}\hat{\Delta}^{-1}.
 \label{sv2}
 \end{equation}
The terms in the right-hand side of Eq. (\ref{sv2}) are the left-off-diagonal matrix, the
diagonal matrix, and the right-off-diagonal matrix, accordingly. Therefore, one obtains a
closed (non-linear) equation for the matrix $\hat T_L$ that leads to homogeneous in time
statistics of the matrix. Next, we obtain for components of the diagonal matrix
$\partial_t{\hat{\Delta}} \hat{\Delta}^{-1}$ expressions that are random variables with
statistics homogeneous in time. Therefore $\ln\Delta_1$, $\ln\Delta_2$, and $\ln\Delta_3$
(where $\Delta_i$ are eigen values of the matrix $\hat\Delta$) are subjects of the
central limit theorem. Typically, the variables behave linear in time $t$ with
coefficients of the order of $\lambda$. The situation with the matrix $\hat T_R$ is
slightly more complicated since there are the exponential factors in the last term of Eq.
(\ref{sv2}). Therefore some components of the matrix $\hat T_R$ behave exponentially in
time like the factors.

Based on the leading role of the shear term in the expression (\ref{sig}), one obtains
from Eq. (\ref{sv2}) for $\hat T_L$ a hierarchy $\chi\gg\chi_1\gg \chi_2$. Therefore in
the principal approximation in $\lambda/s$ the only component, $\sigma_{yx}$, is relevant
and the equation for the matrix $\hat T_L$ is reduced to a single equation for the
component $\chi$:
 \begin{equation}
 \partial_t{\chi}=s-\chi^2\sigma,
 \label{sss}
 \end{equation}
where $\sigma\equiv\sigma_{yx}$. We conclude that the variable $\chi$ possesses a
homogeneous in time statistics, in accordance with our general expectations. Note that
$\chi\sim s/\gamma\gg1$ as follows from Eq. (\ref{sss}). Keeping the main in $\chi$
contributions to the diagonal terms in Eq. (\ref{sv2}) one obtains $\mathrm{diag}\,
(\partial_t{\hat{\Delta}}\hat{\Delta}^{-1}) =(-\chi \sigma, \chi\sigma, 0)$. Therefore in
this approximation
 \begin{equation}
 \mathrm{diag}\, \Delta = (e^{-\rho},e^{\rho},1), \quad
 \partial_t{\rho}=\chi\sigma.
 \label{e61}
 \end{equation}
If $t\gg\lambda^{-1}$ then typically $\rho\sim\lambda t\gg 1$.

One concludes from the equations for $\zeta_1,\zeta_2,\zeta_3$ following from Eq.
(\ref{sv2}) that at $\lambda t\gg1$ the variable $\zeta_1$ is ``frozen'' at a level of
order unity whereas the variables $\zeta_2,\zeta_3$ grow exponentially, they can be
estimated as $e^\rho$. However, the combination $\zeta_1\zeta_3-\zeta_2$, entering $\hat
T_R^{-1}$, is `frozen'' at a level of order unity as well as $\zeta_1$. Based on the
results obtained for the matrices $\hat T_L$, $\hat\Delta$, and $\hat T_R$ one finds
eigen values of the matrix $\hat W$. In the main approximation in $\lambda/s$ we obtain
$W_1\sim e^\rho$, $W_2\sim 1$, $W_3\sim e^{-\rho}$. The expressions together with Eq.
(\ref{e61}) lead to the relation
 \begin{eqnarray}
 \lambda\equiv \langle\partial_t\rho\rangle
 =\langle \chi \sigma \rangle,
 \label{srr}
 \end{eqnarray}
where averaging is performed over the velocity statistics.

Since the variable $\rho$ is an integral over time of a random quantity with homogeneous
in time statistics, see Eqs. (\ref{sss},\ref{e61}), the $\rho$-statistics possesses some
universal features at $\lambda t\gg1$. Namely, at the condition the probability
distribution function (PDF) of $\rho$ can be written in a self-similar form \cite{Ellis}
 \begin{equation}
 P(\rho)\propto \exp[-tS(\rho/t)],
 \label{gor}
 \end{equation}
where $S$ is the so-called Kramer function (or entropy function). The expression
(\ref{gor}) is a manifestation of PDF's for the so-called intensive variables (see, e.g.,
\cite{Landau}). The expression (\ref{gor}) implies that relative fluctuations of $\rho$
diminish as $t$ grows.

Let us consider moments of the divergence of close Lagrangian trajectories in our random
flow. The equation governing a separation between the trajectories $\Delta\bm R$ is
$\partial_t \Delta R_j= \Sigma_{ji}\Delta R_i$, it can be obtained from Eq.
(\ref{divergence}) by putting $\bm\xi\to 0$. A solution of the equation is $\Delta \bm
R(t)=\hat W \Delta \bm R(0)$. Therefore at $t\gg \lambda^{-1}$ we arrive at the
estimation $\Delta R(t)\sim \Delta R(0) e^\rho$. Then the moments of $\Delta \bm R$ can
be calculated in the saddle-point approximation (justified by the inequality $\lambda
t\gg1$):
 \begin{eqnarray}
 \langle |\Delta\bm R|^n \rangle = \int d\rho\ P(\rho)
 |\Delta\bm R|^n \propto \exp(\lambda_n t),
 \label{lambdad} \\
 \lambda_n=-S(\psi_n)+n \psi_n, \quad \mathrm{where} \quad
 S'(\psi_n)=n.
 \label{lambdan}
 \end{eqnarray}
Thus, the exponents $\lambda_n$ are determined by statistical properties of the
Lagrangian trajectories. Note that the Lyapunov exponent $\lambda$ can be formally
expressed via $\lambda_n$ as $\lambda= d\lambda_n/dn|_{n=0}$.

General statistical properties of the separation $\Delta\bm R$ for the random flow with
strong average shear were established in Ref. \cite{05CKLT}, in context of the single
polymer dynamics in such flow. One expects a strong intermittency of $\Delta R(t)$, it
should be revealed in faster than linear growth of $\lambda_n$ at large $n$, since the
linear law $\lambda_n\propto n$ is characteristic of Gaussian statistics of $\Delta
R(t)$.

 \section{Correlation Functions}
 \label{sec:corre}

To find a time dependence of the magnetic field moments one has to perform an additional
average of the expression (\ref{high}) over space that is equivalent to averaging over
the $\rho$ statistics and statistics of initial magnetic fluctuations. Thus, the $2n$-th
moment of the magnetic field induction is written as
 \begin{equation}
 \left\langle {\bm B}^{2n}(t)\right\rangle=
 \int d\rho\ P(\rho) \left\lfloor {\bm B}^{2n}(t)\right\rfloor .
 \label{mom}
 \end{equation}
In our approximation, $W_1\sim e^\rho$, $W_2\sim 1$ and, therefore, in the diffusionless
regime $B(t)\sim e^\rho {\cal B}_0$, whereas in the diffusion regime one obtains $\lfloor
|\bm B|^2 \rfloor \sim {\cal B}_0^2(l/r_d) e^\rho$, as follows from Eq.
(\ref{diffusive}). Substituting the expressions into Eq. (\ref{mom}) and integrating over
$\rho$ (in the saddle-point approximation) we find $\gamma_n=\lambda_{2n}$ for the
diffusionless regime and $\gamma_n=\lambda_n$ for the diffusion regime. Thus, we related
the dynamo increments (growth rates) introduced by Eq. (\ref{increments}) to the
statistical properties of the flow. Our results can be summarized in terms of the
estimations
 \begin{eqnarray}
 \langle |\bm B(t)|^{2n} \rangle
 \sim \exp(\lambda_{2n}t) {\cal B}_0^{2n},
 \quad t<\lambda^{-1} \ln(l/r_d), \quad
 \label{pairmom} \\
 \langle |\bm B(t)|^{2n} \rangle
 \sim (l/r_d)^n\exp(\lambda_{n}t) {\cal B}_0^{2n},
 \ t>\lambda^{-1} \ln(l/r_d). \
 \nonumber
 \end{eqnarray}

The main contribution to the moments $\left\langle {\bm B}^{2n}(t)\right\rangle$ is
associated with the component $B_x$ of the magnetic induction directed along the velocity
of the shear flow, see Fig. \ref{fig:blobs}. Let us turn to moments of the component
$B_y$ directed along the gradient of the shear flow, $\langle B_y^{2n} \rangle$. The
moments are much smaller than the moments $\left\langle {\bm B}^{2n}(t)\right\rangle$,
the smallness is caused by the strong shear flow. One finds from Eqs.
(\ref{e6},\ref{e61}) that $\lfloor B_x^{2}(t) \rfloor=\chi^2 \lfloor B_y^{2}(t)\rfloor$.
Thus, the variable $\chi$ is a measure of the magnetic field anisotropy, $\chi^{-1}$
determines the tilt angle $\phi$ of the magnetic blobs to the shear velocity, see Fig.
\ref{fig:blobs}. Since the variable $\chi$ possesses homogeneous in time statistics, the
factor $\chi^{-2}$ does not produce a difference in the growth rates, that is the both
moments, $\langle B_x^{2n} \rangle$ and $\langle B_y^{2n} \rangle$, are proportional to
the same exponent $\exp(\gamma_n t)$. However, the prefactors at the exponents are
different. To find the difference in the prefactors is not enough to know statistical
properties of the variable $\rho$ that determine the exponent. Generally, one should know
a mutual probability distribution of the variables $\rho(t)$ and $\chi(t)$ that is quite
complicated object depending on details of the flow dynamics. However, one can establish
an estimation for typical fluctuations $\chi\sim s/\lambda$ that follows from Eqs.
(\ref{sss},\ref{srr}). Therefore, say, $\lfloor B_x^{2}(t) \rfloor\sim (s^2/\lambda^2)
\lfloor B_y^{2}(t)\rfloor$.

There is a question concerning moments of the third component of the magnetic induction,
$\langle B_z^{2n}\rangle$. To analyze their behavior one should take into account the
components of the matrix $\hat T_L$, that we ignored at investigating $B_x$ and $B_y$.
Then we conclude that the time dependence of $\langle B_z^{2n}\rangle$ is characterized
by the same exponents $\exp(\gamma_n t)$ both at the diffusionless and diffusion stages.
As to prefactors, they depend on details of the flow statistics.

 \subsection{Pair Correlations}

Let us consider the simultaneous magnetic field pair correlation function
 \begin{equation}
 F_{ij}(t,\bm r)
 =\langle B_i(t,\bm r_1+\bm r) B_j(t,\bm r_1) \rangle.
 \label{pair}
 \end{equation}
Here angular brackets mean, as previously, averaging over space (that is an integration
over $\bm r_1$ with the inverse volume as a factor). We assume statistical homogeneity in
space of both the velocity and the initial magnetic field fluctuations, that is why the
space average (\ref{pair}) characterizes the magnetic field correlations in the whole
volume. We consider the case $r\ll\eta$, then we can use the smooth flow approximation.

Again, we start from the presentation (\ref{solution}). Then, analogously to the second
moment, the pair correlation function (\ref{pair}) can be written as
$F_{ij}(t)=\left\langle \lfloor W_{ik} {\cal B}_k[\bm R(0)] W'_{jl} {\cal B}_l[\bm
R'(0)]\rfloor \right\rangle$, where the trajectories $\bm R$ and $\bm R'$ terminate at
the points $\bm r_1+\bm r$ and $\bm r_1$, respectively, at time $t$. Then one obtains
 \begin{equation}
 F_{ij}(t)=\left\langle
 \lfloor W_{ik}(t) W_{jl}(t)
 {\cal F}_{kl}[\Delta\bm R(0)]
 \rfloor\right\rangle,
 \label{pairback1}
 \end{equation}
where ${\cal F}_{ij}$ is the initial (at $t=0$) pair correlation function of the magnetic
field fluctuations and $\Delta\bm R= \bm R- \bm R'$. The correlation length $l$ of ${\cal
F}$ is smaller than $\eta$, that is why we can consider $|\Delta\bm R|<\eta$. Then both
evolution matrices in the expression (\ref{pairback1}) can be taken at the same point
$\bm R$. Averaging in Eq. (\ref{pairback1}) can be treated as averaging over the velocity
statistics.

The difference $\Delta\bm R$ satisfies the same equation (\ref{divergence}) provided
$|\Delta\bm R|\ll \eta$. However, now we are interested in the solution with the final
condition $\Delta\bm R = \bm r$. Such solution is written as
 \begin{eqnarray}
 \Delta\bm R(t')=
 \hat W(t') \hat W^{-1}(t) \bm r
 \qquad \qquad \nonumber \\
 -\hat W(t') \int_{t'}^t dt_1
 \hat W^{-1} (t_1)[\bm\xi(t_1)-\bm\xi'(t_1)],
 \label{difference1}
 \end{eqnarray}
instead of Eq. (\ref{difference}). We immediately conclude from Eq. (\ref{difference1})
that the pair correlation function coincides with the second moment if $r\lesssim r_d$.
Therefore further we examine the case $r\gg r_d$ where the second term in Eq.
(\ref{difference1}) is negligible and we find
 \begin{equation}
 \Delta\bm R(0)= \hat W^{-1}(t) \bm r.
 \label{difference2}
 \end{equation}
To be more precise, the expression (\ref{difference2}) is correct provided $y\sim r\gg
r_d$, that is implied below.

There are two different regimes for the pair correlation function. If $t<\lambda^{-1}
\ln(l/r)$ then $\Delta R(0)$ is typically less than $l$, the regime exists if $r\ll l$.
In this case the two Lagrangian trajectories, $\bm R$ and $\bm R'$ remains typically
within the correlation radius $l$ at $t=0$ and the behavior of the expression
(\ref{pairback1}) is insensitive to the separation $r$. Therefore the pair correlation
function $F_{ij}$ practically coincides with the single-point average $\langle B_iB_j
\rangle$ in this regime and ,consequently, its time dependence is determined by the
increment $\gamma=\lambda_2$.

If $t>\lambda^{-1} \ln(l/r)$ then $\Delta \bm R(0)$ is typically larger than $l$ and only
rare events where $|\Delta \bm R(0)|<l$ contribute to the correlation function. Using the
representation $\hat{W}=\hat{T}_L\hat{\Delta}\hat{T}_R$ one obtains from Eq.
(\ref{difference2}) an expression $\Delta R(0)\approx e^{\rho} (r_x-\chi r_y)$, where
$r_x$ and $r_y$ are coordinates of the separation $\bm r$. A probability that the
quantity is less or of the order of $l$ is estimated as $e^{-\rho} l/r$ (provided
$r_x\sim r_y \sim r$), that is an interval of values of $\chi$ where $\Delta R(0)<l$.
Therefore $B_iB_j\sim {\cal B}_0^2 e^{\rho} l/r$ and, consequently, $F(t)\sim {\cal
B}_0^2 \exp(\lambda_1 t) l/r$.

Let us collect the obtained results
 \begin{eqnarray}
 F(t)\sim {\cal B}_0^2 \exp(\lambda_2 t),
 \quad t<\lambda^{-1} \ln(l/r)
 \nonumber \\
 F(t)\sim {\cal B}_0^2 \exp(\lambda_1 t) l/r,
 \quad t>\lambda^{-1} \ln(l/r),
 \label{paircf}
 \end{eqnarray}
where the inequality $r_d\ll r \ll l$ is assumed. Thus, the pair correlation function is
governed by the same exponents as the second moment. In addition, we find an
$r$-dependence of the pair correlation function. Note that the expression (\ref{paircf})
turns into the expression (\ref{pairmom}) for the second moment at $r\sim r_d$ as it
should be.

Returning to the expression (\ref{pairback1}) we conclude that a difference between the
pair correlation function $F_{ij}$ and the moments $\langle B_i B_j \rangle$ is solely in
the behavior of $\Delta \bm R$. Therefore relations between components of $F_{ij}$
controlled by the evolution matrices in Eq. (\ref{pairback1}) is the same as for the
moments $\langle B_i B_j \rangle$, say, $F_{yy}\sim (\lambda/s)^2 F_{xx}$.

 \subsection{Mellin transform}

It is instructive to examine the Mellin transform of the pair correlation function, the
analysis reveals its scaling properties. We define the Mellin transform as
 \begin{equation}
 \tilde F(t,k)=\int\limits_0^\infty
 \frac{d r}{r} \left(\frac{r}{l}\right)^{-ik} F(t,r),
 \label{Mellin}
 \end{equation}
where a direction of the radius-vector $\bm r$ is implied to be fixed. Due to smoothness
of the velocity field, different harmonics $\tilde F(t,k)$ evolve independently, being
represented as a sum of exponents characterizing different structures of $\tilde F_{ij}$.
At times $t\gg \lambda^{-1}$ only the principal exponent survives, that is $\tilde
F(k,\varphi)\propto \exp[\gamma(k)t]$. In order to return back to the real space one
should perform the inverse Mellin transform
 \begin{equation}
 F(t,r)=\int\limits_{-\infty}^{\infty}\frac{dk}{2\pi}
 \exp\left[-ik\ln\frac{l}{r}+\gamma(k)t\right]\tilde {\cal F}(k).
 \label{inverse}
 \end{equation}
The quantity $\tilde{\cal F}(k)$ is determined by the initial magnetic field
fluctuations, correlated on the scale $l$. That is why we incorporated the quantity into
the relations (\ref{Mellin},\ref{inverse}).

Some words about analytical properties of $\tilde {\cal F}(k)$. We assume that at $r>l$
the initial pair correlation function ${\cal F}(r)$ decreases fast as $r$ grows. Then the
integral (\ref{Mellin}) converges if $\mathrm{Im}\ k>0$. Therefore $\tilde {\cal F}(k)$
is analytical in the upper $k$-semiplane. Besides, the integral diverges (at small $r$)
as $k\to 0$. Therefore singularities of $\tilde {\cal F}(k)$ lie in the lower
$k$-semiplane, starting from the point $k=0$. A character of the singularities depends on
analytical properties of the initial function ${\cal F}(r)$. If it is analytical in $r$
then one expects that $\tilde {\cal F}(k)$ has a series of poles along the lower
imaginary semiaxis, and the first one lies at $k=0$. Note that in accordance with general
rules the integration contour in Eq. (\ref{inverse}) should go above the first singular
point $k=0$.

Now we can draw some general conclusions taking into account that $\gamma(k)\sim
\lambda$. If $\ln(l/r)>\lambda t$ then the integral (\ref{inverse}) is determined by a
narrow vicinity of the point $k=0$. Then $F(t,r)\propto \exp[\gamma(0)t]$ and we identify
$\gamma(0)$ and $\lambda_2$. If $\ln(l/r)<\lambda t$ then the integral (\ref{inverse})
can be calculated in the saddle-point approximation. To find the saddle point one should
shift the integration contour into the upper semiplane to reach the saddle-point
$k=iq_\star$, where $q_\star$ determines the minimal value of $\gamma(iq)$ at $q>0$.
Indeed, the increment $\gamma(iq)$ is real and therefore the point $k=iq_\star$ is a
solution of the equation $d\gamma/dk=0$ giving an extremum of the exponent in Eq.
(\ref{inverse}). Then $F(t,r)\propto (l/r)^{q_\star} \exp[\gamma(iq_\star)t]$, and we
identify $\gamma(iq_\star)$ with $\lambda_1$, see Eq. (\ref{paircf}). Note that in
accordance with the asymptotic law (\ref{paircf}) $q_\star$ should be equal to unity,
$q_\star=1$.

 \subsection{High Order Correlations}

Here we consider higher order correlation functions of the magnetic field
(\ref{corrfun}). One obtains expressions like Eq. (\ref{difference1}) for separations
$\Delta \bm R$ between the points $\bm R_1(0), \dots ,\bm R_{2n}(0)$ that are needed to
calculate $F_{2n}$ in accordance with Eq. (\ref{solution}). If $\lambda t < \ln
(l/|\Delta \bm r|)$ for all separations between the points $\bm r_1, \dots ,\bm r_{2n}$
then all separations $\Delta\bm R(0)$ are less than $l$. In this situation we arrive at
the same expression $F_{2n}\sim {\cal B}_0^{2n} \langle e^{2n\rho}\rangle$ as for the
$2n$-th moment, and we conclude that $F_{2n}\propto \exp(\lambda_{2n}t)$, see Eq.
(\ref{pairmom}).

Now we turn to the case where $\lambda t >\ln (l/r)$ where all separations $\Delta\bm r$
are assumed to be of the same order. Let us consider first the geometry where all the
points $\bm r_1, \dots, \bm r_{2n}$ lie on a line, that is all vectors $\bm r_\alpha-\bm
r_\beta$ have the same directions, and one can write $\Delta \bm r\sim \bm r$, where $\bm
r$ is one of the separations. Then one arrives at the estimation $|\Delta \bm R|\sim
e^{\rho} (r_x-\chi r_y)$, similar to the pair correlation function. Thus, one obtains the
same probability $\sim e^{-\rho} l/r$ that the separations $|\Delta \bm R(0)|\lesssim l$.
Then we find $F_{2n}\sim {\cal B}_0^{2n} \langle e^{(2n-1)\rho}\rangle l/r$, where the
factor $e^{2n\rho}$ originates from the product of the matrices $\hat W$, appearing in
accordance with the expression (\ref{solution}). Averaging the expression, one obtains
$F_{2n}(t)\sim {\cal B}_0^{2n} \exp(\lambda_{2n-1} t) l/r$. Let us stress that the growth
rates here are different from the growth rates of the corresponding moments.

However, the above expression is correct only if $t<\lambda^{-1} \ln(l/r_d)$. For larger
$t$ the diffusion contributions to the differences $\Delta_\alpha \bm R$ become relevant,
see Eq. (\ref{difference1}). Then, manipulating by $\chi$, it is possible to make less
than $l$ only one difference among $\Delta \bm R(0)$. After that all other differences
acquire typically values of the order of $r_d e^\rho$, and a probability that a
difference is smaller than $l$ is estimated as $(l/r_d)e^{-\rho}$. Thus we conclude
$F_{2n}\sim\langle (l/r) (l/r_d)^{n-1} e^{n\rho}\rangle \propto \exp(\lambda_n t)$.

The same results (upto combinatoric factors) are true for a collinear geometry where the
set $\bm r_1, \dots, \bm r_{2n}$ is separated into $n$ pairs with parallel vectors $\bm
r_\alpha-\bm r_\beta$ characterizing the pairs. Then the corresponding differences
$\Delta\bm R$ behave like previously and the same arguments are correct. Let summarize
our results for the collinear geometry
 \begin{eqnarray}
 F_{2n}(t) \sim {\cal B}_0^{2n} \exp(\lambda_{2n}t) ,
 \quad t<\lambda^{-1} \ln(l/r); \quad
 \nonumber \\
 F_{2n}(t)\sim {\cal B}_0^{2n} \exp(\lambda_{2n-1} t) l/r , \qquad
 \label{highcf} \\
 \lambda^{-1} \ln(l/r)<t<\lambda^{-1} \ln(l/r_d);  \qquad
 \nonumber \\
 F_{2n}(t) \sim {\cal B}_0^{2n} (l/r)(l/r_d)^{n-1}\exp(\lambda_{n}t),
 \ t>\lambda^{-1} \ln(l/r_d). \
 \nonumber
 \end{eqnarray}
It is a generalization of the expressions (\ref{paircf}) for the pair correlation
function. Note also, that, as for the pair correlation function, in the collinear
geometry the estimate $B_y\sim (\lambda/s) B_x$ determines relations between different
components of $F_{2n}$ whereas $z$-components need a separate investigation.

If the collinear geometry is destroyed, then it is impossible to put all the separations
$|\Delta\bm R(0)|$ inside the scale $l$ at any $\chi$, if $t>\lambda^{-1}\ln(l/r)$. In
this situation a behavior of the correlation function $F_{2n}$ is non-universal being
sensitive to details of the initial space distribution of ${\cal B}$. Any case, a value
of $F_{2n}$ in the non-collinear geometry is much less than in the collinear one. The
situation resembles one realized for the randomly advected passive scalar on scales
larger than the pumping length \cite{99BFLL}. We conclude that at $t>\lambda^{-1}
\ln(l/r)$ correlations of the magnetic field are concentrated near collinear geometries,
decaying away from the geometries. The decaying length is estimated as $l e^{-\lambda t}$
at $t<\lambda^{-1} \ln(l/r_d)$ and as $r_d$ at $t>\lambda^{-1} \ln(l/r_d)$.

 \section{Short Correlated Flow}
 \label{sec:short}

Here we consider a strong steady shear flow complemented by a random component shortly
correlated in time. The case admits an analytical solution and can be, therefore, used to
varify our general assertions and predictions. In addition, the case is naturally
realized since the strong shear destroys correlations of the random component, that is
why we expect that the short-correlated case is widely spread in real flows.

In the short-correlated case the matrix of the velocity gradients $\hat\sigma$ describing
the random component of the flow has to be treated as white noise, that is a variable
$\delta$-correlated in time. In the isotropic case one obtains the following tensorial
structure
 \begin{eqnarray}
 \langle \sigma_{ik}(t_1\!)\sigma_{jn}(t_2\!)\rangle
 \!=\!D(4\delta_{ij}\delta_{kn}\!-\!\delta_{ik}\delta_{jn}
 \!-\!\delta_{in}\delta_{jk})\delta(t_1\!-\!t_2), \quad
 \label{isotropic}
 \end{eqnarray}
where the factor $D$ characterizes strength of the random flow and the numerical factor
is introduced as in Ref. \cite{95CFKL}. However, as we already argued, the only relevant
component of the random velocity gradients matrix in the case $\lambda\ll s$ (that is a
manifestation of the random flow weakness) is $\sigma\equiv \sigma_{yx}$. We characterize
its statistical properties by the expression
 \begin{equation}
 \langle \sigma(t_1) \sigma(t_2) \rangle = 4D \delta(t_1-t_2),
 \label{short}
 \end{equation}
formally coinciding with Eq. (\ref{isotropic}) for the $yx$-component. Other components
of $\hat\sigma$ can have correlation functions different from the expression
(\ref{isotropic}). The random component can be treated as weaker than the steady shear
flow provided $D\ll s$.

Statistical properties of the separation between close Lagrangian trajectories $\Delta\bm
R(t)$ in the random smooth flow with strong shear component in the short-correlated case
are investigated in Ref. \cite{07Tur} (in the context of polymer dynamics). We present
here results obtained in the paper without a derivation. Note that our variable $\chi$ is
related to the tilt angle $\phi$ of Ref. \cite{07Tur} as $\chi=\cot\phi$, $\chi\approx
\phi^{-1}$ in the case of small tilt angles, we are interested in. An expression for the
Lyapunov exponent found in Ref. \cite{07Tur} is
 \begin{equation}
 \lambda=\frac{\sqrt\pi\, 3^{1/3}}{\Gamma(1/6)} D^{1/3}s^{2/3}.
 \label{fok3}
 \end{equation}
Thus, the condition $s\gg D$ guarantees the inequality $\lambda \ll s$, indeed. Note also
that $\lambda\gg D$ and that $\lambda\to0$ as $D\to0$. The last property is a natural
consequence of zero value of the Lyapunov exponent for a purely shear flow.

For the short-correlated case, it is possible to find the exponents $\lambda_n$
characterizing growth rates of the moments of $\Delta\bm R(t)$, see Eq. (\ref{lambdad}),
if $n\gg1$. Then a saddle-point (instanton) approximation in the functional space can be
used, see Ref. \cite{96FKLM}, that leads to
 \begin{equation}
 \gamma_n=\frac{3}{2^{5/3}}n^{4/3} D^{1/3} s^{2/3}
 \sim \lambda n^{4/3},
 \label{gammam}
 \end{equation}
see Ref. \cite{07Tur}. The non-linear dependence of $\lambda_n$ on $n$, $\lambda_n\propto
n^{4/3}$, signals about strong intermittency of the flow. Note, that in our anisotropic
case the growth rates $\lambda_n$ increase as $n$ grows slower then in the isotropic
case, where for the short-correlated flow $\lambda_n\propto n^{2}$, see Ref.
\cite{99CFKV}.

 \subsection{Pair Correlation Function}

Next, we examine the two-point simultaneous correlation function (\ref{pair}) for the
short-correlated in time velocity field. In the case it is possible to derive a closed
equation for the correlation function (see, e.g., Ref. \cite{01FGV}). We study here the
function at scales that are much larger than the diffusion scale $r_d$ (but much smaller
than the velocity correlation length $\eta$). Then it is possible to neglect diffusion
effects and we omit all terms with the noise $\bm\xi$ in subsequent relations.

Let us shortly explain a derivation of the equation. First of all, it follows from the
definition (\ref{evolution}) that
 \begin{eqnarray}
 \hat W(t) = \mathrm T\exp(\hat \Xi)
 \hat W(t-\tau), \quad
 \hat \Xi =\int_{t-\tau}^t dt'\ \hat\Sigma,
 \label{decom}
 \end{eqnarray}
where $\tau$ is an arbitrary time (less than $t$). We choose $\tau$ to be much smaller
than $\lambda^{-1}$ but much larger than the velocity correlation time (it is possible
for the short-correlated flow). Then $\hat \Xi$ is a small factor, though the two factors
in Eq. (\ref{decom}) can be treated as statistically independent. Substituting the
expression (\ref{decom}) into Eq. (\ref{pairback1}), expanding the result into a series
over $\hat \Xi$ (upto the second order) and averaging the result inside the time interval
$(t-\tau,t)$ in accordance with Eq. (\ref{short}), one obtains a variation of $F_{ij}$ at
passing from $t-\tau$ to $t$. Since the variation is small, it can be rewritten in terms
of a differential equation.

Assuming the isotropic correlation function of the fluctuations (\ref{isotropic}) one
obtains the following equation
 \begin{eqnarray}
 \partial_t F_{ij}
 =-2sy \partial_xF_{ij}+sF_{iy}\delta_{jx}
 +sF_{yj}\delta_{ix}+
 \nonumber \\
 +4D\left[\delta_{ij} F_{kk}-\frac{1}{2}F_{ij}
 -r_k\partial_j F_{ik}-r_k\partial_i F_{kj}+ \right . \quad
 \nonumber \\
 \left . +\frac{1}{2}(\bm r\nabla)F_{ij}
 +\frac{1}{2}r^2\nabla^2 F_{ij}
 -\frac{1}{4}r_{m}r_{n}
 \partial_{m}\partial_{n} F_{ij}\right]. \qquad
 \label{main_eq}
 \end{eqnarray}
In the absence of the shear (at $s=0$) the system of equations (\ref{main_eq}) leads to a
closed equation for the trace of the correlation function $H= F_{kk}$:
 \begin{eqnarray}
 \partial_t H
 =D\left(10H+6r\partial_r H
 +r^2\partial_r^2 H\right).
 \nonumber
 \end{eqnarray}
The equation coincides with one presented in Ref. \cite{Kaz} (for the case of the scaling
exponent $\xi=2$ and zero forcing).

Now we rule out irrelevant terms in Eq. (\ref{main_eq}) using the following properties:
$s$ is much larger than $D$, characteristic value of $x$ is much larger then that of $y$
and, respectively, $\partial_y \gg\partial_x$. Then we can keep in Eq. (\ref{main_eq})
solely the terms originating from $\sigma_{yx}$. The resulting equation leads to a closed
system of equations for the three components, $F_{xx}$, $F_{xy}$, $F_{yy}$, of the pair
correlation function:
 \begin{eqnarray}
 \partial_{t} F_{xx}=-2sy \partial_x
 F_{xx}+2sF_{xy}+4Dx^2\partial^2_{y}F_{xx},
 \qquad \qquad \label{system_fgh} \\
 \partial_{t}
 F_{xy}=-2sy \partial_x
 F_{xy}+sF_{yy}
 +2Dx^2\partial^2_{y}F_{xy}
 -4Dx\partial_y F_{xx},
 \nonumber \\
 \partial_{t} F_{yy}=-2sy \partial_x
 F_{yy}+2Dx^2\partial^2_{y}F_{yy}
 -8Dx\partial_y F_{xy}+DF_{xx}.
 \nonumber
 \end{eqnarray}
Further, we use the dimensionless time $T={(8Ds^2)^{1/3}}t$ and introduce the
designations $f=F_{xx}$, $g=(s/D)^{1/3} F_{xy}$ and $h=(s/D)^{2/3} F_{yy}$.

Let us investigate a special case of coinciding points. At $\bm r=0$ all terms with
derivatives drop from the equations (\ref{system_fgh}) and they take the form
 \begin{equation}
 \partial_T \left(
  \begin{array}{c}
    f \\
    g \\
    h \\
  \end{array}
 \right)=\left(
          \begin{array}{ccc}
            0 & 1 & 0 \\
            0 & 0 & 1/2 \\
            2 & 0 & 0 \\
          \end{array}
        \right)\left(
  \begin{array}{c}
    f \\
    g \\
    h \\
  \end{array}
 \right).
 \label{r=0_system}
 \end{equation}
A growing solution of the solution is
 \begin{equation}
 \left(
 \begin{array}{c}
    f \\
    g \\
    h \\
 \end{array}
 \right)_{r=0}\propto\left(
 \begin{array}{c}
    1 \\
    1 \\
    2 \\
  \end{array}
 \right)e^T .
 \label{"r=0"}
 \end{equation}
The behavior corresponds to the increment $\gamma=(8Ds^2)^{1/3}$ of the magnetic field
second moment. From the other hand, the case corresponds to small $r$ that is to the
condition $\ln(l/r)>\lambda t$. Therefore $\gamma=\lambda_2$, and we conclude that
$\lambda_2=(8Ds^2)^{1/3}$ in our case.

It is convenient to pass into the ``polar coordinates'' $\varrho,\varphi$ in the shear
plane: $x= \varrho \cos \varphi$, $y =(D/s)^{1/3} \varrho \sin \varphi$. Then we perform
the Mellin transform $f,g,h\to \tilde f,\tilde g,\tilde h$ in terms of $\varrho$ and
derive the equations for $\tilde f,\tilde g,\tilde h$ from the system (\ref{system_fgh}).
In terms of the quantity $q=-ik$ the equations are written as
 \begin{eqnarray}
 \partial_T \tilde f=\tilde f^{\prime\prime} \cos^4\varphi+\tilde f^{\prime}
 \left[\sin^2\varphi-2(q+1)\sin\varphi\cos^3\varphi\right]
 \nonumber \\
 +\tilde f\left[q\sin\varphi\cos\varphi-q\cos^4\varphi
 +(q^2+q)\sin^2\varphi\cos^2\varphi\right]+\tilde g,
 \nonumber \\
 \partial_t \tilde g=\tilde g^{\prime\prime} \cos^4\varphi
 +\tilde g^{\prime}\left[\sin^2\varphi-2(q+1)\sin\varphi\cos^3\varphi\right]+
 \nonumber \\
 +\tilde g\left[q\sin\varphi\cos\varphi-q\cos^4\varphi+(q^2+q)\sin^2\varphi\cos^2\varphi\right]
 \nonumber \\
 -2\cos^2\varphi\ \tilde f^{\prime}+2q\sin\varphi\cos\varphi\ \tilde f +\frac{1}{2}\tilde h,
 \nonumber \\
 \partial_t \tilde h=\tilde h^{\prime\prime} \cos^4\varphi
 +\tilde h^{\prime}\left[\sin^2\varphi-2(q+1)\sin\varphi\cos^3\varphi\right]+
 \nonumber \\
 +\tilde h\left[q\sin\varphi\cos\varphi-q\cos^4\varphi
 +(q^2+q)\sin^2\varphi\cos^2\varphi\right]
 \nonumber \\
 -4\cos^2\varphi\ \tilde g^{\prime}+4q\sin\varphi\cos\varphi\ \tilde g+2\tilde f, \qquad
 \label{tr_system}
 \end{eqnarray}
where prime designates a derivative over the angle $\varphi$.

Next, we study the time evolution of the system (\ref{tr_system}) numerically for
different real values of $q$ (so that $k=iq$ is imaginary) using implicit difference
scheme on the interval $(-\pi/2,\pi/2)$ for $\varphi$ with the periodic boundary
conditions. We have chosen as initial conditions for $f,g,h$ the same Gaussian functions
centered near $\varphi=0$ and with width of order unity. Then we extract the principal
increment $\gamma(iq)=(8Ds^2)^{1/3} c(q)$ dominating the behavior of the system at
$T\gg1$. The dimensionless quantity $c$ was extracted as
 \begin{eqnarray}
 c=\frac{1}{T}\ln\frac{f(T_0+T)}{f(T_0)},
 \nonumber
 \end{eqnarray}
where $T_0+T$ is chosen to be large enough (near $30$) and $T_0$ is introduced to exclude
an influence of an initial transient process, we have chosen $T_0=T$.

\begin{figure}[t]
 \includegraphics[width=3.5in,angle=0]{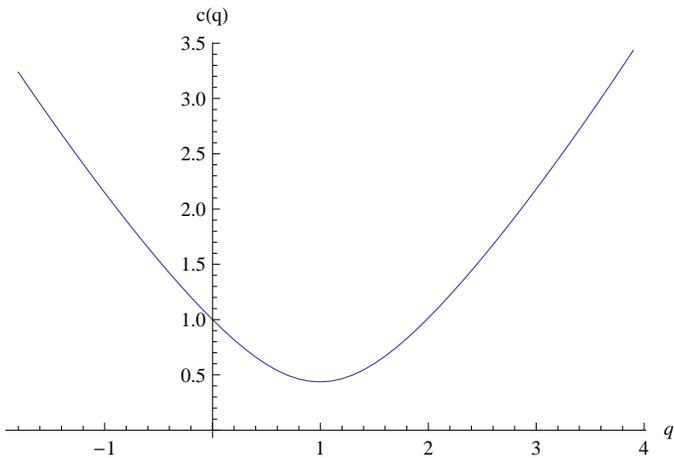}
 \caption{Increment of the Mellin transform of the pair correlation function on the imaginary axis.}
 \label{fig:cp}
 \end{figure}

The quantity $c$ is plotted as a function of $q$ in Fig. \ref{fig:cp}. It turned out to
be positive everywhere, having a minimum at $q=1$, $c(1)=0.435$. The value $q=1$ is in
accordance with Eq. (\ref{paircf}) and the general analysis given in Section
\ref{sec:corre}. As we argued there, the minimum value of $\gamma(iq)$ determines
$\lambda_1$, that is $\lambda_1=c(q_\star) (8Ds^2)^{1/3}$. The value of $c$ at $q=0$ is
$c=1$, in accordance with Eq. (\ref{"r=0"}) and general arguments given in Section
\ref{sec:corre}. Thus, the obtained results confirm our general assertions.

 \section{Discussion}
 \label{sec:discussion}

We have analyzed the kinematic dynamo stage when the small-scale fluctuations of the
magnetic field grow in a steady shear flow complemented by relatively weak random
velocity fluctuations. The weakness is characterized by the inequality $s\gg\lambda$
where $s$ is the shear rate and $\lambda$ is the Lyapunov exponent of the flow. Universal
features we are established are revealed at times $t\gg \lambda^{-1}$. The shear makes
the flow strongly anisotropic, which, paradoxically, simplifies the analysis of the
dynamo phenomenon since a single component of the random velocity gradient appears to be
relevant. We analyzed the situation where the correlation length $l$ of initial magnetic
field fluctuations is less than the velocity correlation length $\eta$ (that is the
Kolmogorov length for developed turbulence). Probably, the smallness of $l$ is not
crucial for our scheme since small scales of the magnetic field distribution in space are
inevitably produced by the hydrodynamic motion.

Let us underline that in the main approximation in $\lambda/s$ our problem is reduced to
a purely two-dimensional velocity field (with components along the shear velocity and
along its gradient). We have proved an existence of the dynamo in this case (that is the
exponential growth of the magnetic field moments). The result obviously contradicts the
statement of Refs. \cite{Zeldovich1,Zeldovich2,Zeldovich3} (Zeldovich theorem) that there
cannot be the magnetic dynamo in two-dimensional flows. We assert that this statement is
wrong and the error is in ignoring the third component $B_3$ of the magnetic induction
(perpendicular to the velocity plane). The third component satisfies the passive scalar
equation and, consequently, decays exponentially. However, $B_3$ cannot be ignored in the
divergentless condition $\nabla\bm B=0$ since the characteristic scale of the magnetic
field along direction of its growth increases faster than the magnetic field itself. One
can check that all the terms in $\nabla\bm B=\partial_x B_x+\partial_y B_y+\partial_z
B_z$ decay with the same exponent, and therefore the condition $\nabla\bm B=0$ leads to
an effectively divergent in-plane magnetic field. The dynamo effect is not forbidden for
such field. A detailed analysis of the discrepancy will be published elsewhere. An
existence of the dynamo effect for two-dimensional flows is a subject of numerical
verification.

For our general conclusions, we do not specify statistical properties of the random flow
exploring only its smoothness at scales less than the velocity correlation length $\eta$.
Then it is possible to relate the kinematic growth rates $\gamma_n$ of the magnetic
field, see Eq. (\ref{increments}), to intrinsic characteristics of the flow
characterizing divergence of close Lagrangian trajectories, see Eq. (\ref{lambdad}). We
find that $\gamma_n=\lambda_{2n}$ in the diffusionless regime and $\gamma_n=\lambda_n$ in
the diffusion regime. We also related the anisotropy degree of the magnetic field to the
same intrinsic characteristics of the flow. Thus, the main features of the magnetic field
statistics (including its intermittency) are dictated by the flow statistics. Note that
our general scheme can be applied without essential modifications to the statistically
isotropic flows or to random flows with other types of anisotropy.

We established main features of the magnetic field correlation functions. The pair
correlation function behaves like the second moment at small separations $r$, and
increase with the growth rate, characteristic of the diffusion regime, at larger $r$, it
is proportional to $1/r$ there. As to higher order correlation functions, the situation
is more complicated. As small time $t$ they behave as corresponding moments. However, at
larger time, $t>\lambda^{-1}\ln(l/r)$, correlations are peaked near the collinear
geometry (where $2n$ points are separated into $n$ pairs with parallel separations) and
there is an intermediate asymptotic regime when the correlation functions grow with rates
that do not coincide with the growth rates of the moments. Then, at times $t>\lambda^{-1}
\ln(l/r_d)$ the correlation function grows with the same exponent, as the corresponding
moment in the diffusion regime. The scaling behavior of the correlation functions in the
collinear regime is $\propto 1/r$. Correlations decay fast at deviation from the
collinear geometry. That reflects a complicated spacial structure of the magnetic field
that is strongly correlated for special geometries produced by affine geometrical
transformations from initial magnetic fluctuations.

Our general assertions can be verified by solving the model with short correlated in time
fluctuating component, that admits a series of analytical results. The non-linear
$n$-dependence of the growth rates $\gamma_n$, $\gamma_n\propto n^{4/3}$, at large $n$
signals about strong intermittency of the magnetic field. Therefore only rare events
concentrated in a restricted fraction of space contribute to high moments of the magnetic
field. We analyzed in detail the pair correlation function of the magnetic field for the
model, the analysis confirms all our general assertions, including scaling behavior in
different regimes.

The ideology and the analytical approach developed in our work can be expanded to the
dynamics of polymer solutions possessing elasticity that is described similarly to the
magnetic field. Moving along these lines, we hope to clarify some aspects of the
so-called elastic turbulence \cite{00GS,01GSa,01GSb} that are still not explained.

\acknowledgments

We thank A. Chernykh, M. Chertkov, G. Falkovich, and S. S. Vergeles for helpful
discussions and remarks. The work was partially supported by RFBR under grant no.
09-02-01346-a and by MES of RF under FTP ``Kadry".

\end{document}